\documentclass[sigconf]{acmart}
\usepackage{multirow}

\AtBeginDocument{%
  \providecommand\BibTeX{{%
    \normalfont B\kern-0.5em{\scshape i\kern-0.25em b}\kern-0.8em\TeX}}}

\setcopyright{acmlicensed}
\copyrightyear{2018}
\acmYear{2018}
\acmDOI{XXXXXXX.XXXXXXX}

\begin{document}

\title{Effective Two-Stage Knowledge Transfer for Multi-Entity Cross-Domain Recommendation}


\author{Jianyu Guan}
\authornote{Both authors contributed equally to this research.}
\affiliation{%
  \institution{Alibaba Group}
  \city{Hangzhou}
  \country{China}}
\email{guanjianyu.gjy@alibaba-inc.com}

\author{Zongming Yin}
\authornotemark[1]
\affiliation{%
  \institution{Alibaba Group}
  \city{Hangzhou}
  \country{China}}
\email{mocun.yzm@alibaba-inc.com}

\author{Tianyi Zhang}
\affiliation{%
  \institution{Alibaba Group}
  \city{Hangzhou}
  \country{China}}
\email{zty325975@alibaba-inc.com}

\author{Leihui Chen}
\affiliation{%
  \institution{Alibaba Group}
  \city{Hangzhou}
  \country{China}}
\email{leihui.clh@alibaba-inc.com}

\author{Yin Zhang}
\affiliation{%
  \institution{Alibaba Group}
  \city{Hangzhou}
  \country{China}}
\email{jianyang.zy@alibaba-inc.com}

\author{Fei Huang}
\affiliation{%
  \institution{Alibaba Group}
  \city{Hangzhou}
  \country{China}}
\email{huangfei.hf@alibaba-inc.com}

\author{Shuguang Han}
\authornote{Shuguang Han is the corresponding author.}
\affiliation{%
  \institution{Alibaba Group}
  \city{Hangzhou}
  \country{China}}
\email{shuguang.sh@alibaba-inc.com}

\author{Jufeng Chen}
\affiliation{%
  \institution{Alibaba Group}
  \city{Hangzhou}
  \country{China}}
\email{jufeng.cjf@alibaba-inc.com}


\renewcommand{\shortauthors}{Guan and Yin, et al.}

\begin{abstract}
In recent years, the recommendation content on e-commerce platforms has become increasingly rich --- a single user feed may contain multiple entities, such as selling products,  short videos, and content posts. To deal with the multi-entity cross-domain recommendation problem, an intuitive solution is to adopt the shared-network-based architecture for joint training. The underlying idea is to transfer the knowledge from one type of entity (source entity) to another (target entity). 
However, different from the conventional same-entity cross-domain recommendation, multi-entity knowledge transfer encounters several important challenges: (1) data distributions of the source entity and target entity are naturally different, making the shared-network-based joint training susceptible to the negative transfer issue, (2) the corresponding feature schema of each entity is not exactly aligned (e.g., price is an essential feature for selling product while missing for content posts), making the existing approaches no longer appropriate. Recent researchers have also experimented with the pre-training and fine-tuning paradigm. Again, they only take into account the scenarios with the same entity type and feature schemas, which is inappropriate in our case.
To this end, we design a pre-training \& fine-tuning based Multi-entity Knowledge Transfer framework called MKT. MKT utilizes a multi-entity pre-training module to extract transferable knowledge across different entities. In particular, a feature alignment module is first applied to scale and align different feature schemas. Afterward, a couple of knowledge extractors are employed to extract the common and independent knowledge. In the end, the extracted common knowledge is adopted for target entity model training. 
Through extensive offline and online experiments on public and industrial datasets, we demonstrated the superiority of MKT over multiple State-Of-The-Art methods. MKT has also been deployed for content post recommendations on our production system.
\end{abstract}






\maketitle

\section{Introduction}
\begin{figure}[htbp]
  \centering
  \includegraphics[width=0.7\linewidth]{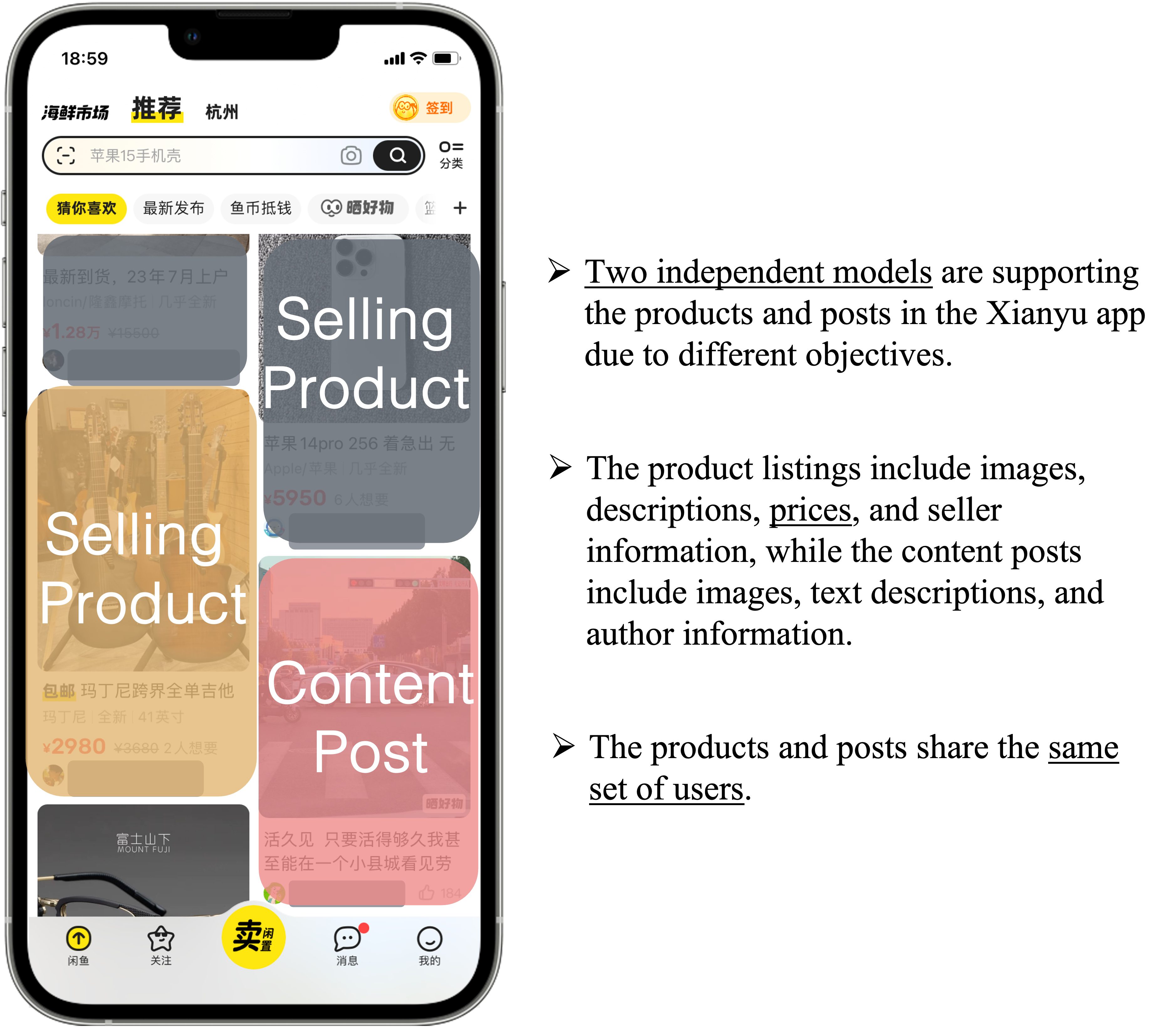}
  \caption{\label{fig:xianyu}Homepage recommendation of Xianyu APP, which consists of a mixture of selling products and content postings.}
\end{figure}

In recent years, the recommendation content on the majority of e-commerce platforms has become increasingly rich -- a single user feed may contain multiple entities, such as selling products, short videos, and content posts. Take Xianyu App\footnote{Xianyu is the largest online flea marketplace in China. It allows every consumer to post and sell their second-hand products on the platform.} for example, its homepage feed consists of both the selling products and content posts (consumers may sometimes post product reviews, unboxing videos, shopping experiences, and et al., and such information is useful for other users with the same purchase needs). As illustrated by Figure~\ref{fig:xianyu}, our production system offers a small proportion of traffic for content posts. To improve the recommendation experience, we focus on building an effective recommendation algorithm for Xianyu content posts in this paper. Despite being focused on Xianyu in this paper, we believe that the proposed approach can be easily applied to other e-commerce platforms as long as they face the same multi-entity recommendation problem.

Due to the limited exposure, building an independent model~\cite{cheng2016wide,guo2017deepfm,zhou2018deep,zhou2019deep,ouyang2019deep} solely on the scarce impression data of content posts can be inferior in model performance~\cite{zhang2019deep}.  An intuitive solution is to utilize abundant user interaction information from the other entity (source domain)~\cite{zhu2022personalized,khan2017cross}, i.e. the selling products for Xianyu, and assist in the training of the content post (target domain) recommendation algorithm. Despite being promising, an effective transfer of knowledge across different entities faces several important challenges: (1) data distributions of the source entity and target entity are naturally different, making the shared-network-based joint training susceptible to the negative transfer issue, and (2) the feature schemas of different entities are not exactly aligned. For instance, price is an important feature for products while it is unavailable for content posts. This makes the conventional cross-domain recommendation approaches~\cite{zhang2022keep,liu2023continual} inappropriate.
To the best of our knowledge, current research on cross-domain recommendation can be broadly divided into two groups: the \textbf{joint training}~\cite{elkahky2015multi,li2020ddtcdr,hu2018conet,Ma2018ModelingTR,Tang2020ProgressiveLE,chang2023pepnet,zhang2023collaborative,zhao2020catn,ouyang2020minet,li2021dual} approach in which data from source and target domains is trained together within a single stage, and the \textbf{pre-training \& fine-tuning}~\cite{chen2021user,zhang2022keep,yang2022click,liu2023continual,huan2023samd} approach with the first stage on pre-training the source domain and the second stage on fine-tuning the target domain. Joint training algorithms typically transfer knowledge from the source domain to the target domain by designing shared parameters, most of them only consider the same entity type~\cite{Ma2018ModelingTR,chang2023pepnet,sheng2021one}, while a few studies consider multiple entities~\cite{ouyang2020minet,hu2018conet,li2020ddtcdr}. However, the shared-parameter architecture often encounters the negative transfer~\cite{zhang2021survey} problem during back-propagation. This is more prominent once the training data is highly skewed towards one domain. In our production system, the amount of data for the product entity is 3 times over the content post entity, this may cause the model parameters to be governed by the product entity~\cite{crawshaw2020multi,xie2022multi}. To address negative transfer problem, researchers have tried pre-training and fine-tuning. First, a base model is trained using a lot of source data. Then, the obtained model is adjusted with target domain data, or a new target model is tuned with the knowledge from the obtained model to fit the target distribution better~\cite{zhang2022keep,liu2023continual}. However, they only take into account the scenarios with the same entity type and feature schema.

To address the problem of multi-entity knowledge transfer with heterogeneous feature schema, we propose an effective \textbf{M}ulti-entity \textbf{K}nowledge \textbf{T}ransfer algorithm (MKT for short). MKT follows the pre-training \& fine-tuning paradigm.

During pre-training, MKT extracts useful knowledge from the source entity through training a multi-entity compatible base model. To better accommodate for the second-stage fine-tuning, MKT adopts the mixed data from both source and target entities for pre-training and further develops a Heterogeneous Feature Alignment (HFA) module to align the heterogeneous feature schema of the two entities. In addition, MKT employs one Common Knowledge Extractor (CKE) and two Independent Knowledge Extractor (IKE) to extract the common and independent knowledge between two entities. To better capture the semantic meaning of different knowledge representations, MKT further designs an auxiliary task to enforce the common knowledge and independent knowledge to be dissimilar. The resulting multi-entity base model is then plugged into the target domain model for fine-tuning. MKT performs hierarchical transfer of knowledge with different semantics, including user features, entity-shared features, sequence features and user-entity cross vector, and uses gate to filter the knowledge from the pre-training stage to obtain more suitable parts. This ultimately enhances the prediction performance for the target model (i.e., the content post recommendation model). 

To summarize, the main contributions of our work are as follows:
\begin{itemize}
    \item[--] We propose an MKT framework to deal with the problem of multi-entity knowledge transfer with heterogeneous feature schema. MKT first trains a multi-entity base model with the mixed-domain data, and the resulting model is then applied to the target domain for knowledge transfer.
    
    \item[--] By diving into the multi-entity recommendation problem, we discover several important issues and develop the corresponding solutions. Specifically, we develop a Heterogeneous Feature Alignment (HFA) module to handle different feature schemas from multiple entities and propose an auxiliary task to better extract common knowledge across entities. We perform hierarchical transfer of knowledge with different semantics and use gate to filter them.
    
    \item[--] We verify the effectiveness of MKT on public and industrial datasets and examine the model performance through online A/B testing on the Xianyu platform. Experimental results demonstrate that MKT outperforms the State-Of-The-Art algorithms by a fair margin, resulting in a 4.1\% increase in click-through rate and a 7.1\% increase in user engagement metric in online experiments.
\end{itemize}
\section{Related Work}
The multi-entity recommendation problem described in this paper falls under the research direction of Cross-Domain Recommendation (CDR for short). The CDR research aims to utilize the abundant data from the source domain to enhance the recommendation performance of the target domain~\cite{fernandez2012cross,zang2022survey,zhu2021cross}. Existing cross-domain algorithms can be divided into two groups: the joint training approach and the pre-training \& fine-tuning approach.

\textbf{CDR with Joint Training.} Joint training approaches attempt to transfer knowledge from the source domain to the target domain by designing proper network architectures with shared parameters.

Considering that users are normally consistent across domains, many studies have investigated the ways of propagating user interest to different domains. For instance, MVDNN~\cite{elkahky2015multi} designed a shared user tower and multiple independent item towers, that helped broadcast user knowledge across domains. MiNet~\cite{ouyang2020minet} divided cross-domain user interests into long-term interest and short-term interest, and jointly models three types of user interests with an attention mechanism. However, a pure transfer of user profile knowledge is insufficient, researchers have also examined the ways of knowledge extraction from user-item feature interaction. MMOE~\cite{Ma2018ModelingTR} utilized multiple shared expert networks and domain-specific gating networks to capture the commonality and difference in user-entity interaction information. PLE~\cite{Tang2020ProgressiveLE} added independent expert networks and exploited a progressive routing mechanism to extract shared and task-specific knowledge. CoNet~\cite{hu2018conet} adopted cross-connection units for knowledge transfer across different domains. STAR~\cite{sheng2021one} proposed a star-shaped topology compromising globally shared parameters and domain-specific parameters, and the final domain parameters were computed with the dot product of the two parameters. PEPNet~\cite{chang2023pepnet} designed a dynamic parameter generator that is shared across multiple domains. 

However, joint training algorithms inevitably face gradient conflict problems, and since the number of data samples in the source domain is usually much larger than that in the target domain, the final model is dominated by the source domain; therefore, the final model does not fully adapt to the target domain.

\textbf{CDR with Pre-training \& Fine-tuning.} The pre-training \& fine-tuning approach is another main solution to the cross-domain recommendation problem. Conventional methods~\cite{chen2021user,yang2022click} usually adopted the same model for both pre-training and fine-tuning. That is, we first pre-train on the source domain to obtain a base model and then fine-tune the resulting base model with the data from the target domain. However, as mentioned in previous studies~\cite{he2021analyzing, kumar2022fine}, those methods can easily get stuck in the local optima once the target data distribution is far away from the source data distribution. Therefore, later studies such as KEEP~\cite{zhang2022keep} adopted the knowledge plugging framework, in which the pre-trained knowledge is inserted into the target domain. However, due to the large amount of training data and the corresponding model architecture, the extracted knowledge from KEEP is rather static. CTNet~\cite{liu2023continual} further proposed a continuous knowledge transfer algorithm for dynamic knowledge transfer. The above knowledge plugging paradigm is indeed capable of adapting the final model to the target domain through fine-tuning; however, they can not deal with entities with different feature schemas.

Regardless of the extensive research on CDR, few of them can be directly applied to effectively handle the multi-entity recommendation problem as the underlying feature schemas can be significantly different. Fortunately, our proposed MKT algorithms can deal with the heterogeneous feature schemas with a well-designed HFA module. It also prevents the model from being dominated by the source domain through transferring knowledge from a pre-trained multi-entity model to the target-entity model.
\section{Methodology}
This section introduces the proposed MKT algorithm. We define our research topic as the multi-entity cross-domain recommendation, in which we attempt to utilize rich user interaction data from the source entity to enhance the Click-Through Rate (CTR) for the target entity. Formally, let us denote the source entity data as $(X^s, Y^s)$ and the target entity data as $(X^t, Y^t)$, where $X$ represents the input features and $Y$ represents the click label information. Note that the notations in this paper follow the below rules: the bold uppercase letters define matrices, the bold lowercase letters define vectors, and regular lowercase letters define scalars. All of the vectors are in the form of column vectors.

\subsection{Model Overview}
\label{subsec:model-overview}
\begin{figure*}[htbp]
  \centering
  \includegraphics[width=0.75\linewidth]{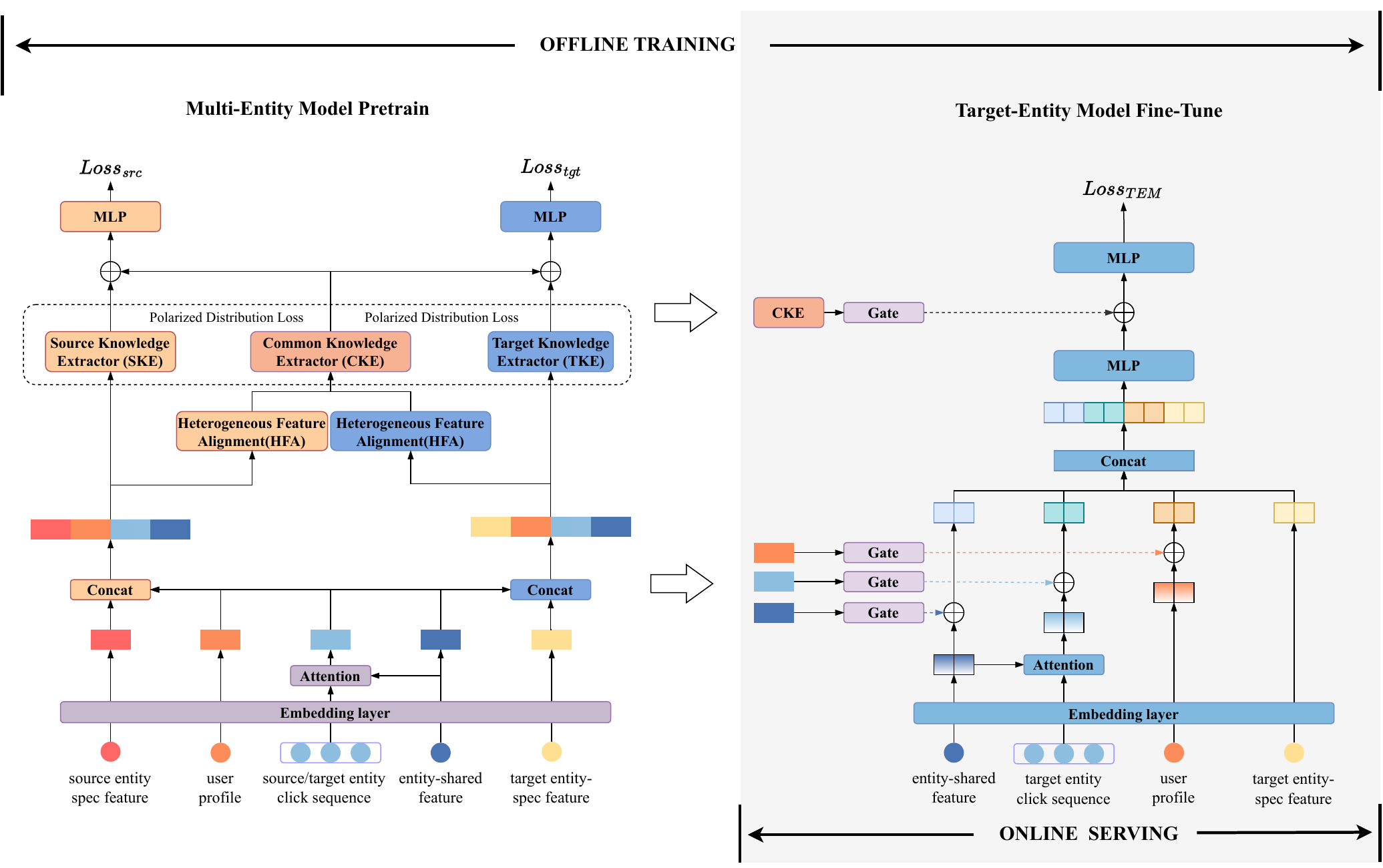}
  \caption{An overview of the model architecture for MKT, which consists of a multi-entity model (left) that is trained on the mixed-domain data, and a target entity model (right) for the target entity recommendation. Modules in the gray color denote the online serving part.}
  \label{fig:model}
\end{figure*}

As illustrated by Figure~\ref{fig:model}, MKT consists of two stages. In the first stage, we utilize both $(X^s, Y^s)$ and $(X^t, Y^t)$ to pre-train a Multi-Entity Model (MEM). In the second stage, we build a Target Entity Model (TEM) and connect it with MEM through joint training. Ultimately, the resulting blended model (see the gray color in Figure~\ref{fig:model}) will be adopted to serve the target entity recommendation. 

Considering the feature difference, MEM develops a Heterogeneous Feature Alignment (HFA) module to scale and align different feature schemas, facilitating the knowledge extraction process in the subsequent Common Knowledge Extractor (CKE) module. To ensure CKE functions as intended, we introduce two independent knowledge extractors to focus on extracting the independent knowledge, and an auxiliary loss to enforce the independent knowledge to be away from the common knowledge. 

In the second stage, the TEM is trained with the frozen low-level feature embeddings and extracted common knowledge of MEM to predict user clicks on the target entity. Instead of directly concatenating the learned representations, we introduce the gate layers to better accommodate the target entity model. The whole process is also visualized in Figure~\ref{fig:model}.

\subsection{Multi-Entity Model Pre-training}
\label{subsec:mem}
In the below sections, we provide more details on pre-training the MEM with mixed source and target entity data. 


\subsubsection{Feature Embedding}
To effectively leverage multi-entity data, we categorize the input features into four groups, as outlined below.

\begin{itemize}
\item[--] \textbf{User profile feature}. Each user has her profile features such as user ID, age group, gender, and city, which represent the static information of a user.

\item[--] \textbf{Multi-entity user behavior sequence}. Since we have two types of entities in Xianyu, there are correspondingly two types of user behavior sequences: one for the source entity (products the user has clicked) and the other for the target entity (content posts the user has clicked). 

\item[--] \textbf{Entity-shared feature}. Features that are the same across the source entity and the target entity, such as the entity catalog, the creator ID, and et al. We refer to those overlapped features as the \textit{entity-shared feature}.

\item[--] \textbf{Entity-specific feature}. This denotes the unique features that only exist in one type of entity. For instance, the product entity has a price feature not present in the content post entity. Similarly, the content entity owns a content style feature, which is not included in the product entity.
\end{itemize}

All of the above features are first transformed into one-hot encodings, and then converted into low-dimensional dense embedding vectors suitable for the deep neural networks. For each data sample in the source domain $\mathbf{x}^s$, we map it into the below embedding vector format $\mathbf{v}^s$, where $\mathbf{v}^s \in \mathbb{R}^{d_s \times 1}$.
\begin{equation}
    \label{equ:1}
    \mathbf{v}^s = [\mathbf{v}_{u}^s||\mathbf{v}_{seq}^s||\mathbf{v}_{shared}^s||\mathbf{v}_{spec}^s]
\end{equation}
\begin{equation}
    \label{equ:2}
    \mathbf{v}_{u}^s=[\mathbf{e}_1||\mathbf{e}_2||...||\mathbf{e}_{n_u}]
\end{equation}
Here, $\mathbf{v}_{u}^s$ represents the embedding vector of user features, $\mathbf{e}$ denotes the embedding vector of one feature, $n_u$ is the number of user features,   $\mathbf{v}_{shared}^s$ and $\mathbf{v}_{spec}^s$ corresponds to the embedding vector of entity-shared features and entity-specific features respectively. $\mathbf{v}_{seq}^s$ is obtained through the attention mechanism. Since attention is not the key design of this paper, we omit the detail in this section. It can be Multi-Head Attention~\cite{vaswani2017attention} or other sequence encoders. Features employed in the attention mechanism are part of entity-shared features, so multiple entities can share parameters in the sequence encoder. The symbol $||$ denotes the vector concatenation operation. Similarly, for target domain sample $\mathbf{x}^t$, we can obtain $\bold{v}^t \in \mathbb{R}^{d_t \times 1}$ in the same manner. 

\subsubsection{Heterogeneous Feature Alignment(HFA)}
\label{subsubsec:hfa}
In practice, entity-specific features significantly affect the model performance, some features are beneficial to the extraction of common knowledge, while others are harmful. And, with heterogeneous feature schemas, we have to design a module that can properly handle the alignment of entity-specific features. To this end, we develop a Heterogeneous Feature Alignment (HFA) structure that is capable of scaling and aligning those entity-specific features. 
\begin{figure}[htbp]
  \centering
  \includegraphics[width=0.8\linewidth]{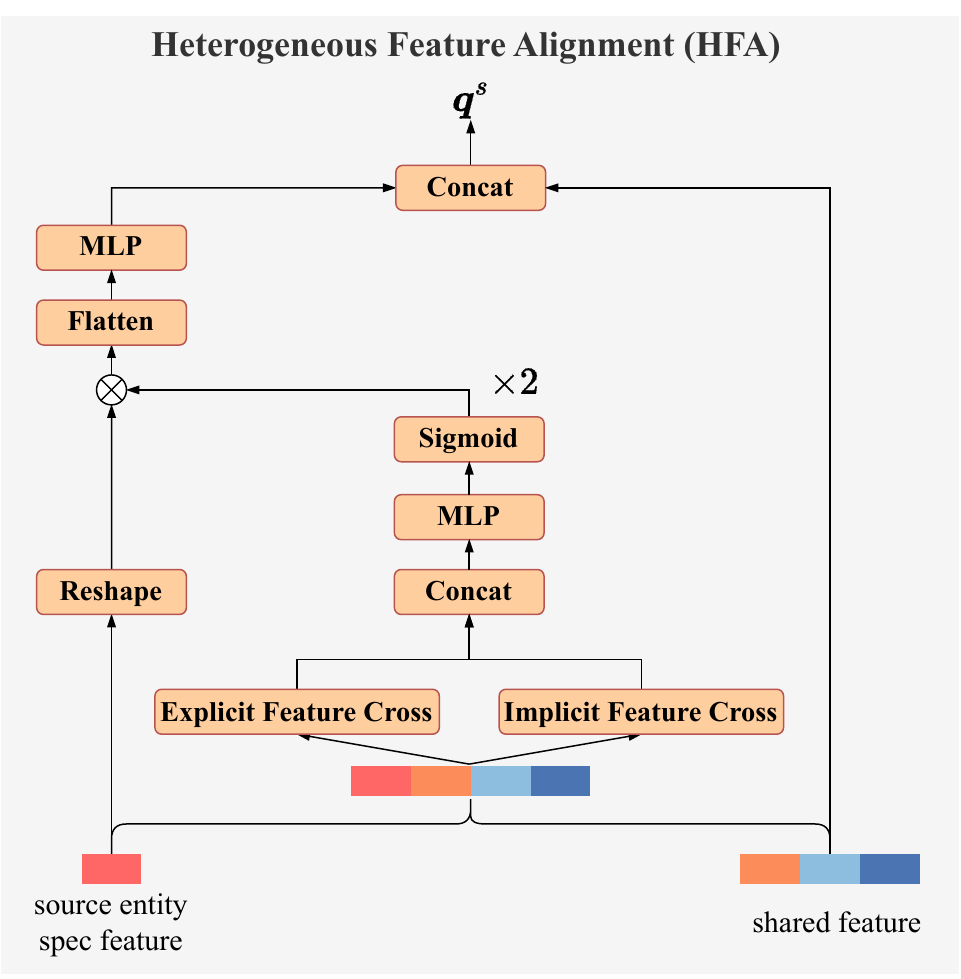}
  \caption{\label{fig:cross}An illustration of the HFA component.}
\end{figure}
HFA first utilizes explicit and implicit feature cross between entity-specific features and shared features (user profile, click sequence, entity-shared feature) to jointly assess the importance of each entity-specific feature. Then, HFA augments those features with the importance score. Finally, HFA maps the entity-specific features (from both source and target entities) into the same dimension for alignment, and then concatenates the resulting vector with the shared features. The whole process can be illustrated as \autoref{fig:cross}. 
Taking source domain samples as an example. For $\mathbf{v}^s_{spec}$, we adopt the Cross network in DCN~\cite{wang2017deep} as the explicit feature cross and exploit multilayer
perceptron(MLP) as the implicit feature cross. Here, we use \textit{ex} and \textit{im} to denote the word \textit{explicit} and \textit{implicit} represently.
\begin{equation}
    \label{equ:5}
    \mathbf{v}^s_{ex} = DCN(\mathbf{v}^s_{spec}), \mathbf{v}^s_{im} = MLP(\mathbf{v}^s_{spec})
\end{equation}
We then concatenate the resulting vectors from the explicit feature cross and implicit feature cross, and exploit an FC layer with a Sigmoid activation function to map them to vectors whose length corresponds to the total number of entity-specific features. This creates the importance score for each feature in the source entity $\mathbf{p}^s\in \mathbb{R}^{n_{spec}^s\times1}$. 
\begin{equation}
    \label{equ:6}
    \bold{p}^s = 2 * Sigmoid(FC(Concat(\mathbf{v}_{ex}^s,\bold{v}^s_{im})))
\end{equation}
Based on the feature importance score, HFA then amplifies important features and down-weights the unimportant ones. In the end, the processed entity-specific features and entity-shared features are concatenated for the use of the next module. The alignment of the source and target entity features is achieved through MLP, which is shown below:
\begin{equation}
    \label{equ:7}
    \bold{V}^s = Reshape(\bold{v}_{spec}^s),\bold{V}^s\in \mathbb{R}^{n_{spec}^s\times d_e}
\end{equation}
\begin{equation}
    \label{equ:8}
    \bold{q}^s = Concat(\bold{v}_{shared}^s,MLP(Flatten(Multiply(\bold{V}^s,\bold{p}^s))))
\end{equation}
where $d_e$ is the dimension of embedding vector of one feature, $Multiply$ represents element-wise multiplication, and $Flatten$ is the operation of flattening a matrix into a column vector. Finally, we obtain output vectors of the source entity and target entity with the same dimension, denoted as $\bold{q}^s \in \mathbb{R}^{d_{HFA}\times1}$ and $\bold{q}^t \in \mathbb{R}^{d_{HFA}\times1}$ ($d_{HFA}$ is the hyper-parameter for HFA).

\subsubsection{Common Knowledge Extractor (CKE)}
After acquiring the aligned entity vectors from HFA, we need to extract the shared knowledge from source domain and target domain. In theory, any network architecture with shared parameters can serve as the common knowledge extractor for MKT, such as MLP or MMoE~\cite{Ma2018ModelingTR}. Here, we opt for the MMoE model structure as our common knowledge extractor, referred to as CKE.

More specifically, we use $f_{CKE}$ to represent the network operations of CKE. Then, we input the output vectors $\mathbf{q}^s$ and $\mathbf{q}^t$ into CKE, as shown in the following formula:
\begin{equation}
    \label{equ:11}
    \bold{g}^s_{com} = f_{CKE}(\bold{q}^s),\bold{g}^t_{com} = f_{CKE}(\bold{q}^t), \bold{g}^s_{com},\bold{g}^t_{com}\in \mathbb{R}^{d_{CKE}}
\end{equation}
where $d_{CKE}$ is the output dimension of the CKE.

\subsubsection{Polarized Distribution Loss (PDL)}
\label{subsubsec:pd}
During joint training, it is hard to measure whether these shared parameters indeed learn the common knowledge, and to what extent they are doing so. To address this issue, we propose an auxiliary task that enforces the Common Knowledge Extractor (CKE) to surely learn the common knowledge. We name the corresponding loss function as Polarized Distribution Loss. Note that along with the CKE, we have also designed independent knowledge extractors for different entities: the Source Knowledge Extractor (SKE) and the Target Knowledge Extractor (TKE). These entity-independent knowledge extractors can be as simple as MLP. For simplicity, we use $f_{SKE}$ and $f_{TKE}$ to denote the knowledge extractors for the source and target entity respectively. Then, we can obtain the independent knowledge $\mathbf{g}^s_{ind}$ and $\mathbf{g}^t_{ind}$ for the source and target entity through the following formula:
\begin{equation}
    \label{equ:12}
    \bold{g}^s_{ind} = f_{SKE}(\bold{v}^s),\bold{g}^t_{ind} = f_{TKE}(\bold{v}^t),\bold{g}^s_{ind},\bold{g}^t_{ind}\in \mathbb{R}^{d_{CKE}}
\end{equation}
We set the output vector dimension of SKE and TKE to be the same as that of CKE. We want the independent and common knowledge to be as different as possible. If the corresponding knowledge extractors capture similar knowledge, the information gain for the target domain would be minimal. So we adopt the PDL to minimize the similarity between them. The cosine function is a commonly adopted similarity function for this purpose. This process can be illustrated by the below formula:
\begin{equation}
    \label{equ:13}
    cosine(\bold{g}^s_{com},\bold{g}^s_{ind}) = \frac{\bold{g}^s_{com}\cdot\bold{g}^s_{ind}}{\Vert \bold{g}^s_{com} \Vert_2\cdot \Vert \bold{g}^s_{ind} \Vert_2}
\end{equation}
$\Vert\cdot \Vert_2$ means the length of a vector in the Euclidean space. With the above notations, the PDL is as follows:
\begin{equation}
    \label{equ:14}
    \begin{split}
        Loss_{pdl} &= Loss_{src\_sim} + Loss_{tgt\_sim}\\
        & = cosine(\bold{g}^s_{com},\bold{g}^s_{ind})+ cosine(\bold{g}^t_{com},\bold{g}^t_{ind})
    \end{split}
\end{equation}
Finally, the MEM loss can be computed as follows:
\begin{equation}
    \label{equ:15}
    \begin{split}
    Loss_{MEM} &= Loss_{src} + Loss_{tgt} + \gamma Loss_{pdl}\\
    &=\sum_{x_i\in X^s}^{N_s} L(y_i^s,f_{MEM}(x_i,\Theta_{MEM}))\\
    &+ \sum_{x_i\in X^t}^{N_t} L(y_i^t,f_{MEM}(x_i,\Theta_{MEM}))+ \gamma Loss_{pdl}
    \end{split}
\end{equation}
where $x_i$ stands for a data sample either from the source or the target domain, $\gamma$ is the hyper-parameter for $Loss_{pdl}$, $f_{MEM}$ represents the operators of MEM, $L$ denotes the loss function, and $\Theta_{MEM}$ indicate the trainable parameters in MEM.

\subsection{Target-Entity Model Fine-tuning}
\label{subsec:tem-finetuning}
\label{subsec:tem}
In the first stage, we train the MEM with mixed data samples from both the source and target entity. However, we notice that the number of data samples for the source entity is far more than that of the target entity. This will result in the MEM being dominated by the source entity and encounters the negative transfer problem~\cite{zhang2021survey}. To overcome this issue, we introduce a second stage fine-tuning process: we connect the TEM with the obtained MEM model from the first stage, and integrate the corresponding knowledge of MEM into the TEM for fine-tuning. By doing so, we can achieve better model performance.

To be more specific, we freeze the parameters of MEM during the second stage. Only target entity samples are required for TEM fine-tuning. To ensure that TEM keeps the entity-specific ability, we only transfer knowledge vectors with the same semantics. These include the user profile feature embedding vector, user behavior sequence vector, entity-shared feature embedding vector, and the user-entity cross knowledge vector extracted by the CKE module.

Considering that the knowledge vectors in the MEM is not entirely applicable to TEM, it is necessary to filter out irrelevant knowledge. This helps to ensure that the resulting knowledge vectors are truly useful for the target domain. To achieve this, we use an adaptive gating structure (such as GLU~\cite{dauphin2017language}) when plugging the knowledge vectors from MEM. For example, the user-entity cross knowledge vector has a gate denoted as $f_{cross\_gate}$, after receiving knowledge from the MEM, we get the final user-entity cross vector for the TEM, and the formulas are as follows,
\begin{equation}
    \label{equ:15}
    \bold{g}_{cross\_final}^{TEM}=\bold{g}_{cross}^{TEM} + f_{cross\_gate}(\bold{g}_{com}^t)
\end{equation}
all of the knowledge vectors such as $\bold{v}_{u\_final}^{TEM}$, $\bold{v}_{seq\_
final}^{TEM}$ and so on can be obtained in the same manner. The overall loss of TEM can be formulated as follows:
\begin{equation}
    \label{equ:17}
    Loss_{TEM} = \sum_{x_i\in X^t}^{N_t}L(y_i^t,f_{TEM}(x_i,\Theta_{TEM}))
\end{equation}
where $x_i\in X^t$ denotes a data sample from the target domain, $f_{TEM}$ represents the operators of the TEM, $L$ denotes the cross-entropy loss adopted in this paper, and $\Theta_{TEM}$ is the trainable parameter for TEM. 
\subsection{Online Deployment}
MKT has been successfully deployed on Xianyu App to serve millions of users daily. The overall system architecture can be shown in \autoref{fig:model}. It is important to note that despite the full MKT architecture being exploited for offline training, we only need the gray part of the MKT for online serving. This is because our model is designed for content post recommendation, in which the modules related to product recommendation are no longer usable. It also allows us to extract the required knowledge with minimal computation cost, which is crucial for online serving.

\section{Experiment}
\subsection{Experiment Setup}
\subsubsection{Datasets}
We conduct our experiments on public datasets and industrial datasets to compare our model and baseline models. The overall statistics are summarized in Table~\ref{tab:dataset_statistics}.

\textbf{Tenrec dataset}~\cite{Tenrec}: This dataset was released by Tencent and is the largest public dataset available for multi-entity cross-domain task that we could find. It is a user behavior log for two different feeds recommendation platforms of Tencent, namely, QQ BOW (QB) and QQ KAN (QK), including click, like, share and follow behaviors. An item in QK/QB can either be a news article or a video. Here, we use QK-video as the source entity and QK-article as the target entity because the number of data samples for QK is far more than that of QB and the same reason for video and article. Consider that QK-article does not contain negative samples for click behavior, we use like behavior type as positive feedback and no-like as negative feedback for QK-video and QK-article. As the dataset presents all interaction behaviors according to the time order, so we use the first 80\% of data for training and the last 20\% for testing.

\textbf{Industrial dataset}: We also construct experimental data with the impression log from our production system. We collect 31 days of user interaction data from the Xianyu homepage recommendation system, 75\% of them are related to selling products and the remaining 25\% is regarding the content posts. In total, we obtain 9.3 billion data samples for the source entity and 3.1 billion data samples for the target entity.  In the below experiments, we use 30 days of data for training and the remaining one day of data for testing.
\begin{table}[htbp]
\caption{Overall statistics of public and industrial datasets. Due to privacy concerns, we are unable to provide the positive sample ratio of the industrial dataset.}
\label{tab:dataset_statistics}
  \begin{tabular}{c|c|c|c|c}
    \toprule
    dataset & \multicolumn{2}{c}{Tenrec} & \multicolumn{2}{c}{Industrial}\\
    \midrule
    domain & video & article & product & post \\
    \midrule
    Users& 5M & 1.3M & 130M & 130M \\
    Items& 1.9M & 0.2M & 290M &  4M \\
    Data samples& 142M& 46M & 9.3B & 3.1B \\
    positive sample ratio & 0.036 & 0.018 & \textbackslash & \textbackslash\\
    \bottomrule
\end{tabular}
\end{table}
\subsubsection{Baselines}
\label{subsec:baseline}
We include both single-domain and cross-domain recommendation algorithms for comparison. The cross-domain approaches are further divided into two categories: the joint training method and the pre-training \& fine-tuning method (\textbf{Finetune} and \textbf{CTNet}).

\textbf{$\rhd$ Single-Domain Recommendation Baselines:}
\begin{itemize}
    \item \textbf{MLP} (Multilayer Perceptron) is the basic deep neural network-based prediction model that includes an embedding layer, a fully connected layer, and an output layer.
    
    \item \textbf{DIN}~\cite{zhou2018deep} (Deep Interest Network) is the widely adopted recommendation algorithm that adaptively learns user interest from historical behavior sequences.
    
    \item \textbf{FiBiNet}~\cite{Huang2019FiBiNETCF} utilizes a Squeeze-and-Excitation network (SENet) to dynamically learn feature importance and employs bilinear interaction to model fine-grained user-item interaction.
\end{itemize}
\textbf{$\rhd$ Cross-Domain Recommendation Baselines:}
\begin{itemize}
    \item \textbf{MMOE}~\cite{Ma2018ModelingTR} (Multi-gate Mixture-of-Experts) leverages a shared mixture-of-experts module and task-specific gating networks for multi-task learning.
    
    \item \textbf{PLE}~\cite{Tang2020ProgressiveLE} (Progressive Layered Extraction) is a multi-task prediction model that compromises both shared and independent expert networks to model shared and task-specific knowledge across tasks.

    \item \textbf{STAR}~\cite{sheng2021one} designs a star-shaped topology to model the shared and independent knowledge of different domains.
    
    \item \textbf{CoNet}~\cite{hu2018conet} (Collaborative Cross Network) employs cross-connection units for knowledge transfer across domains.
    
    \item \textbf{MiNet}~\cite{ouyang2020minet} (Mixed Interest Network) jointly models three types of user interest: 1) long-term user interest across domains, 2) short-term user interest from the source domain, and 3) short-term interest in the target domain.
    
    \item The vallina \textbf{Finetune} method pre-trains a base model ONLY on the source domain data and fine-tunes the resulting model on the target domain data.
    
    \item \textbf{CTNet}~\cite{liu2023continual} follows the pre-training \& and fine-tuning framework, which performs click-through rate prediction under the continual transfer learning setting. It is capable of transferring knowledge from a time-evolving source domain to a time-evolving target domain.
\end{itemize}

\subsubsection{Implementation Details}
Embedding dimensions are set to 8 for all of the sparse features. For the above-mentioned single-domain models and the TEM (see Section~\ref{subsec:tem}), the number of layers for the Deep Neural Network is set to 4, with the number of hidden units in each layer set to [1024, 512, 64, 1] for the industrial dataset and [128, 64, 16, 1] for the public dataset. As for the joint training approaches and the MEM (see Section~\ref{subsec:mem}), we employ a two-layer shared network, with a configuration of hidden units as [1024, 512] for the industrial dataset and [128, 64] for the public dataset. The shared layers are then followed by domain-specific towers for each domain, with each tower consisting of two layers with hidden units [64, 1] for the industrial dataset and [16, 1] for the public dataset. For MMOE, we set the number of expert networks to 3. Since CTNet can not deal with multiple entities, we can only transfer the embedding of user and entity-shared features during fine-tuning. We set $\gamma$ to 0.01 for two datasets. The experiments are conducted multiple times and the mean value is taken as the model performance.

\subsubsection{Evaluation Metrics}
We employ the standard \textbf{AUC} metric (Area Under Receiver Operating Characteristic Curve). A larger AUC implies a better ranking ability. Note that in the production dataset, we further compute the Group AUC (\textbf{GAUC}) to measure the goodness of intra-user ranking ability, which has shown to be more consistent with the online performance~\cite{zhou2018deep,sheng2021one}. GAUC is also the top-line metric in the production system. It can be calculated with Equation~\ref{eqn:gauc}, in which $U$ represents the number of users, $\#\textrm{impression}(u)$ denotes the number of impressions for the $u$-th user, and $\textrm{AUC}_u$ is the AUC computed only using the samples from the $u$-th user. In practice, a lift of 0.1\% GAUC metric (in absolute value difference) often corresponds to an increase of 1\% CTR.
\begin{equation}
\label{eqn:gauc}
\begin{aligned}
    \textrm{GAUC} = \frac{\sum_{u=1}^U \# \textrm{impressions}(u) \times \textrm{AUC}_u}{\sum_{u=1}^U \# \textrm{impressions}(u)}
\end{aligned}
\end{equation}
\subsection{Model Performance}
\subsubsection{Overall Evaluation}

Table~\ref{tab:baselines} provides the model performance for all of the baseline models and our proposed Multi-entity Knowledge Transfer (MKT) model. We discover that single-domain recommendation algorithms are generally inferior to the cross-domain methods, indicating that the data from different domains indeed helps better model user interest.

Firstly, joint-training models generally outperform single-domain models. This superiority arises from joint-training’s ability to leverage common knowledge across multiple domains and enhance the model’s generalization capability. For instance, architectures like MMoE (Mixture-of-Experts) utilize shared expert networks and gating mechanisms to facilitate cross-domain knowledge sharing, resulting in improved recommendation performance across various domains. In contrast, single-domain models rely solely on data from a single domain, it is difficult to achieve good results in target domain with sparse data.

Secondly, the pre-training \& fine-tuning methods further enhances performance beyond joint-training. This improvement is primarily attributed to its two-stage training strategy. During the pre-training stage, the model is trained on a great deal of source data, enabling it to learn universal knowledge representations. Subsequently, during the fine-tuning stage, the model is specifically optimized on target domain data, fine-tuning parameters to align with the target domain’s characteristics. This method not only preserves source domain knowledge but also effectively adapts to the target domain’s specific requirements, mitigating information conflicts and trade-offs that may arise in joint-training.

Finally, our proposed MKT model is further superior to the pre-training \& fine-tuning methods on both AUC and GAUC metrics. This come from two aspects: (1) MKT performs pre-training on the mixed data of source domain and target domain, uses HFA to amplify the entity-spec features suitable for knowledge extracting, aligns different feature schemas, and uses PDL to extract cross-domain common knowledge more efficiently. (2) During the fine-tuning stage, MKT performs hierarchical transfer of knowledge with different semantics, including user features, entity-shared features, sequence features and user-entity cross vector, and uses gate to filter the knowledge from the pre-training stage to obtain more suitable parts.

\begin{table}[htbp]
\caption{A comparison of model performance across different baseline models and the proposed MKT algorithm.}
  \begin{tabular}{cccccc}
    \toprule
    \multicolumn{2}{c}{dataset}&\multicolumn{2}{c}{Tenrec}  &\multicolumn{2}{c}{Industrial} \\
    \midrule
    \multicolumn{2}{c}{Model}&AUC &GAUC &AUC & GAUC  \\
    \midrule
    &MLP&0.9337 & 0.6393 & 0.7315 &  0.5683 \\
    Single-domain&DIN& 0.9381 & 0.6441 & 0.7335 &  0.5761 \\
    &FiBiNet& 0.9374 & 0.6427 & 0.7345 &  0.5824 \\
    \midrule
    &MMoE& 0.9402 & 0.6461 & 0.7367 &  0.5856 \\
 & PLE& 0.9393 & 0.6473 & 0.7363 & 0.5837 \\
 Joint-training& STAR&0.9382 & 0.6422 & 0.7362 & 0.5845 \\
 & CoNet& 0.9368 & 0.6413 & 0.7357 & 0.5820  \\
 & MiNet& 0.9420 & 0.6499 & 0.7370 & 0.5863 \\
 \midrule
 Pre-training& Finetune& 0.9456 & 0.6520 & 0.7372 & 0.5897 \\
 \& & CTNet& 0.9439 & 0.6492 & 0.7380 & 0.5872 \\
 Fine-tuning&  \textbf{MKT} & \textbf{0.9524} & \textbf{0.6583} & \textbf{0.7398} & \textbf{0.5958} \\
 \bottomrule
\end{tabular}
\label{tab:baselines}
\end{table}

\begin{table}[htbp]
\caption{\label{tab:performance_on_user_groups} Model performance on different user groups of the industrial dataset.}
  \begin{tabular}{clclll}
    \toprule
    
    User&Model&\multicolumn{2}{c}{AUC}&\multicolumn{2}{c}{GAUC}\\
 Behaviors& & value& Improv.& value&Improv.\\
    \midrule
    &FiBiNet&  0.7316& -&   0.5659& -\\
    0-10&Finetune&  0.7352&     0.50\%&   0.5773& 2.0\%\\
    &MKT&  0.7379& 0.87\%&   0.5815& 2.8\%\\
    \midrule
    &FiBiNet&  0.7322& -&   0.5883& -\\
 10-30& Finetune&  0.7328& 0.08\%&  0.5976& 1.6\%\\
 & MKT&  0.7354& 0.45\%&  0.6019& 2.3\%\\
 \midrule
 & FiBiNet&  0.7283&-&  0.5973&-\\
 >30& Finetune&  0.7278&0.03\%&  0.6035&1.0\%\\
 & MKT&  0.7310&0.37\%&  0.6074&1.7\%\\
 \bottomrule
\end{tabular}
\end{table}

\begin{table*}[htbp]
\centering
\caption{\label{tab:ablastions} Model performance on different ablation studies.}
  \begin{tabular}{ccccccccc}
    \toprule
    dataset & \multicolumn{4}{c}{Tenrec} & \multicolumn{4}{c}{Industrial}\\
    \midrule
    \multicolumn{1}{c}{\multirow{2}*{Ablations}}&\multicolumn{2}{c}{AUC}&\multicolumn{2}{c}{GAUC} &\multicolumn{2}{c}{AUC}&\multicolumn{2}{c}{GAUC}\\
 & value& Imporv.& value&Improv. & value& Imporv.& value&Improv.\\
    \midrule
    w/o MEM Pretrain & 0.9381 &-1.50\% & 0.6441&-2.15\% &  0.7345 &-0.71\% &  0.5824  & -1.21\%\\
    w/o TEM Finetune & 0.9447 &-0.81\% &0.6503 &-1.21\% &  0.7374 &-0.33\% &   0.5884 &-1.26\%\\
    w/o PDL & 0.9505&-0.19\% & 0.6569& -0.21\%& 0.7390 &-0.11\%& 0.5937 &-0.35\%\\
    w/o HFA &0.9477 &-0.49\% & 0.6522&-0.92\% & 0.7386 &-0.16\%& 0.5906 &-0.88\%\\
    Full MKT&\textbf{0.9524}& - & \textbf{0.6583} & -&\textbf{0.7398} &-&   \textbf{0.5958} &-\\
    \bottomrule
\end{tabular}
\end{table*}

\subsubsection{Model Performance on Different User Groups}
To further understand the behavior of the proposed MKT model, we split users of the industrial dataset into three groups based on their activity levels (which are measured by the number of clicks over the past month). Here, we select FiBiNet and Finetune for comparison. FiBiNet is the best-performing single-domain recommendation model. Finetune falls under the same pre-training \& fine-tuning recommendation paradigm as MKT, and it is also our production baseline.

As shown in Table~\ref{tab:performance_on_user_groups}, the Finetune model outperforms FiBiNet across all three user groups, indicating that the cross-domain approach is superior to the single-domain method. Meanwhile, we see more improvement for users with few behaviors, whereas there is less improvement for users with more behaviors, suggesting that the extracted knowledge from the source entity is more beneficial to long-tail users. When comparing MKT to Finetune, we can reach very similar conclusions. This again demonstrates the effectiveness and robustness of the MKT algorithm.

\subsubsection{Online A/B Testing}
We successfully deployed MKT in our production environment to serve the main traffic of Xianyu APP Homepage recommendation since October of 2024. In this section, we provide online results by comparing MKT with our production baseline. Our production baseline follows the \textbf{Finetune} method in Section~\ref{subsec:baseline}. We conduct an online A/B testing between MKT and our production baseline from October 1 to October 7, each with 5\% randomly assigned traffic. Two key business metrics used for evaluation include the Click-Through Rate (CTR) and User Engagement Rate (UER)~\footnote{UER is one of our top-line business metrics that measures whether a user has liked/commented/collected on a target item.}. During the time of A/B testing, we have observed an increase of CTR metric by +4.13\%, and UER Metric by +7.07\%, which demonstrates the considerable business value of MKT.

\subsection{Ablation Study}
In this section, we conduct a set of ablation studies to further understand the importance of each component in MKT. Particularly, we examine four modules: the MEM pre-training (see Section~\ref{subsec:mem}), the TEM fine-tuning (see Section~\ref{subsec:tem-finetuning}), the HFA (see Section~\ref{subsubsec:hfa}), and the PDL (see Section~\ref{subsubsec:pd}). 

According to Table~\ref{tab:ablastions}, removing any of the four modules can lead to degenerated model performances by a fair margin. By dropping the MEM or TEM module, both AUC and GAUC metrics decrease dramatically, indicating the great importance of them. MKT without the PDL module leads to a drop of 0.11\% and 0.35\% in the AUC and GAUC metrics on industrial dataset, and a drop of 0.19\% and 0.21\% in the AUC and GAUC metrics on public dataset, respectively. Similarly, after removing HFA, model performance drops by 0.16\% in AUC and 0.88\% in GAUC on industrial dataset. The main reason is that HFA helps prescreen feature importance, which eventually ensures more effective knowledge extraction in later modules.

\section{CONCLUSION}
In this paper, we propose a pre-training \& fine-tuning based multi-entity knowledge transfer framework called MKT. MKT can deal with different feature schemas across multiple entities while effectively transferring knowledge from the source entity to the target entity. MKT utilizes a multi-entity pre-training module to extract transferable knowledge across different entities. The heterogeneous feature alignment (HFA) module is designed to scale and align different feature schemas. The polarized distribution loss is used to facilitate an effective extraction of common knowledge. In the end, hierarchical transfer of the extracted knowledge is adopted for training a target entity model. Experiments conducted on public and industry datasets demonstrate MKT outperforms state-of-the-art cross-domain recommendation algorithms. MKT has been deployed in Xianyu App, bringing a lift of +4.13\% on CTR and +7.07\% on UER. 

\bibliographystyle{ACM-Reference-Format}
\bibliography{sample-base}

\end{document}